\documentclass[reprint,
 amsmath,amssymb,
 aps,
]{revtex4-2}

\usepackage{graphicx}
\usepackage{dcolumn}
\usepackage{bm}
\usepackage{xcolor}

\begin{document}


\title{V-Waves: Spatio-temporally induced group-velocity dispersion in free space}

\author{Layton A. Hall$^{1}$}
\author{Ayman F. Abouraddy$^{1}$}
\affiliation{$^{1}$CREOL, The College of Optics \& Photonics, University of Central~Florida, Orlando, FL 32816, USA}

\begin{abstract}
Introducing precise spatio-temporal structure into a pulsed optical field can lead to remarkable changes with its free propagation. `Space-time' (ST) wave packets, for example, propagate rigidly at a tunable group velocity in free space by inculcating a one-to-one relationship between the \textit{axial} wave numbers and the temporal frequencies. Here we introduce a new class of ST wave packets that we call V-waves (so named because of their characteristically V-shaped spatio-temporal spectrum) in which a linear one-to-one relationship is introduced between the temporal frequencies and the \textit{transverse} wave numbers (or spatial frequencies). We confirm experimentally that V-waves experience anomalous group velocity dispersion in free space, all the while maintaining the group velocity fixed at the speed of light in vacuum. Extremely large values of group velocity dispersion can be easily realized, which are not accessible with traditional optical materials or photonic structures. Moreover, V-waves are the unique optical wave packets whose diffraction and dispersion lengths are intrinsically equal by virtue of the spatio-temporal structure of the field itself. These results are of interest for optical signal processing and nonlinear optics.
\end{abstract}

\maketitle

\section{Introduction}

In free space, plane-wave optical pulses do \textit{not} experience dispersive spreading \cite{SalehBook07}. However, \textit{spatially} structuring the field can lead to temporal changes in the pulse profile. For example, spatially confined waveguide modes experience group velocity dispersion (GVD), which can be tailored by varying the waveguide cross section size and shape \cite{mogilevtsev98OL,Ibanescu04PRL,Turner06OE}. Even freely propagating pulsed fields can experience changes in their temporal profile after introducing a spatial structure due to coupling between the spatial and temporal degrees of freedom (DoFs) \cite{Martinez1984JOSA,Zhu05OE,Dorrer19IEEE,Jolly20JO}. This gives rise to the following question: can spatio-temporal couplings introduced into a pulsed beam or wave packet enable engineering its GVD in free space?

Multiple approaches towards spatio-temporally structuring optical wave packets have been recently explored with the aim of controlling their \textit{group velocity} in free space. One embodiment is the so-called `flying focus' whose group velocity is varied by introducing longitudinal chromatism, but the pulse spectrum concomitantly evolves along the propagation axis \cite{SaintMarie17Optica,Froula18NP,Jolly20OE}. We focus here on another embodiment termed `space-time' (ST) wave packets \cite{Kondakci16OE,Parker16OE,Kondakci17NP,Yessenov19OPN}, which are a family of pulsed optical beams endowed with a precise association between the spatial and temporal DoFs \cite{Donnelly93ProcRSLA,Longhi04OE,Saari04PRE,Kondakci19OL,Yessenov19PRA,Yessenov19OE}, early examples of which include focus-wave modes (FWMs) \cite{Brittingham83JAP,Reivelt00JOSAA,Reivelt02PRE} and X-waves \cite{Lu92IEEEa,Saari97PRL}, among other examples \cite{Reivelt03arxiv,Turunen10PO,FigueroaBook14,Shen21PRR}. These freely propagating wave packets can be endowed with a variety of useful characteristics, such as propagation invariance \cite{Kondakci18PRL,Bhaduri18OE,Bhaduri19OL,Yessenov20NC}, tunable group velocities \cite{Salo01JOA,Recami03IEEEJSTQE,Valtna07OC,Zamboni08PRA,Wong17ACSP2,Efremidis17OL,Porras17OL,PorrasPRA18,Kondakci19NC,Bhaduri19Optica}, self-healing \cite{Kondakci18OL}, and anomalous refraction \cite{Bhaduri20NP}, and can even be realized using incoherent fields \cite{Yessenov19Optica,Yessenov19OL}. Furthermore, new modes of interaction are enabled by these wave packets with photonic devices, such as planar cavities \cite{Shabahang17SR,Shiri20OL,Shiri20APLP}, waveguides \cite{Zamboni01PRE,Zamboni02PRE,Zamboni03PRE,Shiri20NC,Kibler21PRL}, and surface plasmon polaritons \cite{Schepler20ACSP}.

The hallmark of ST wave packets is a one-to-one relationship between the temporal frequency and the \textit{axial} wave number \cite{Donnelly93ProcRSLA}. For example, propagation invariance necessitates a linear relationship between the temporal frequency and the \textit{axial} wave number \cite{Donnelly93ProcRSLA,Kondakci17NP,Turunen10PO}, which entails a group velocity in free space that differs from $c$ (the speed of light in vacuum) \cite{Yessenov19PRA}, with the sole exception of Brittingham's FWM \cite{Brittingham83JAP}. Furthermore, deviating from this linear dispersion relationship can inculcate GVD into the field \cite{Liu98JMO,Hu02JOSAA,Lu03JOSAA} and can enable the realization of arbitrary dispersion profiles \cite{Yessenov21arxiv}. 

Here we introduce theoretically and verify experimentally a new ST wave packet that we call a V-wave because of its V-shaped spatio-temporal spectrum. A V-wave is endowed with a unique set of attributes stemming from its spatio-temporal spectrum, whereupon the temporal frequency is linearly related to the \textit{transverse} wave number. This condition entails a nonlinear relationship between the temporal frequency and axial wave number, thus leading to anomalous GVD experienced by V-waves in free space. The magnitude of this GVD is easily tuned to extremely large values that are not available in non-resonant optical materials or photonic structures. Therefore, V-waves provide resonant-like GVD values at any wavelength \cite{Yessenov20OSAC} without losses or limitations on bandwidth \cite{Kondakci18OE}. Unlike propagation-invariant ST wave packets, V-waves travel in free space at a group velocity of $c$. The unique spatio-temporal spectral structure introduces a `Janus'-like spectral phase (Janus is the two-faced Roman god) that governs the propagation dynamics: when expanded in terms of spatial frequencies, the spectral phase represents diffraction in space; and when expanded in terms of temporal frequencies, the spectral phase corresponds to dispersive propagation in time. Consequently, the diffraction and dispersion lengths of V-waves are \textit{intrinsically equal}. Through simulations and measurements, we confirm these unique attributes of V-waves.

\begin{figure*}[ht!]
\begin{center}
\includegraphics[width=17.8cm]{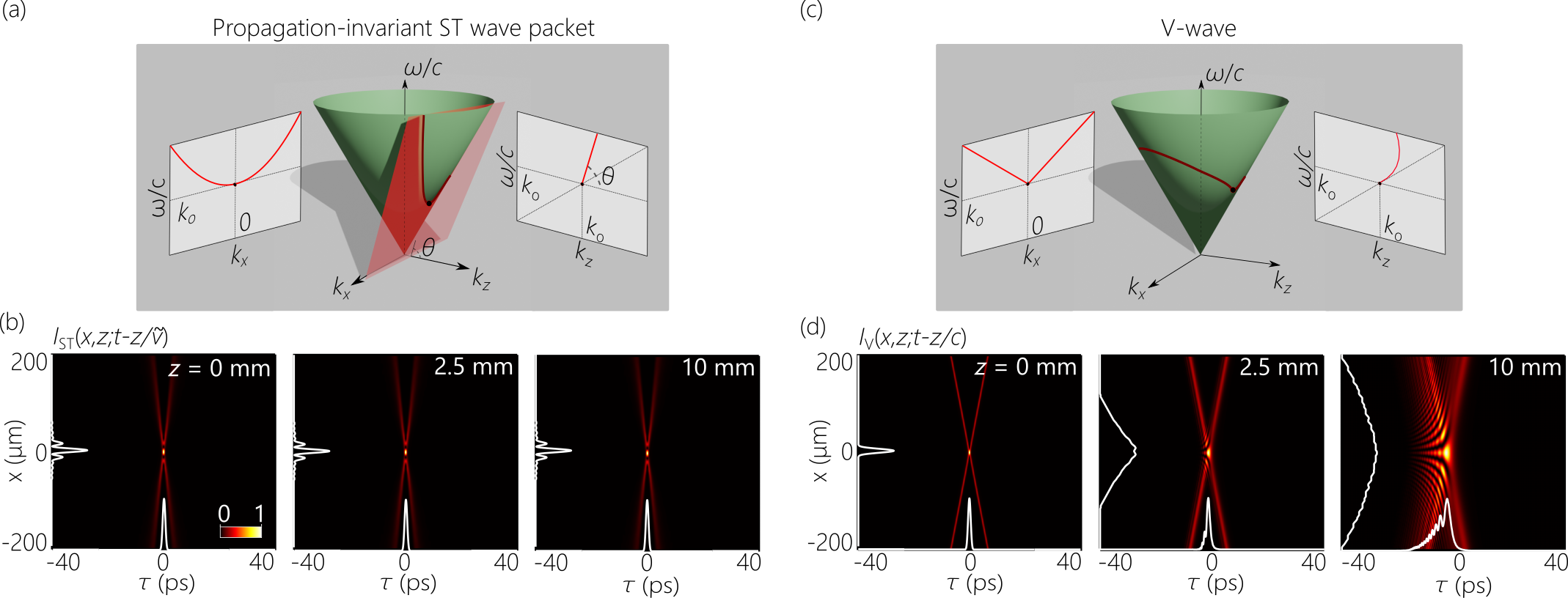} 
\end{center}
\caption{The concept of V-waves. (a) The spectral support domain of a propagation-invariant ST wave packet in $(k_{x},k_{z},\tfrac{\omega}{c})$ space lies at the intersection of the light-cone with a plane that is parallel to the $k_{x}$-axis and makes an angle $\theta$ with respect to the $k_{z}$-axis. The spectral projection onto the $(k_{z},\tfrac{\omega}{c})$-plane is a straight line and onto the $(k_{x},\tfrac{\omega}{c})$-plane a conic section. (b) The propagation-invariant spatio-temporal profile $I_{\mathrm{ST}}(x,z;t)$ at $z\!=\!0$, 2.5, and 10~mm for $\theta\!=\!49^{\circ}$, $\lambda_{\mathrm{o}}\!=\!800$~nm, and bandwidth $\Delta\lambda\!=\!2$~nm. (c) The spectral support domain of a V-wave with positive-valued $\alpha$ on the surface of the light-cone. The spectral projection onto the $(k_{x},\tfrac{\omega}{c})$-plane is V-shaped and onto the $(k_{z},\tfrac{\omega}{c})$-plane is a conic section that is tangential to the light-line. (d) The spatio-temporal profile of a V-wave with $\alpha\!=\!0.1$, $\lambda_{\mathrm{o}}\!=\!800$~nm, and $\Delta\lambda\!=\!2$~nm at the same axial positions in (b). In (b) and (d) we plot in white $I(x\!=\!0,z;\tau)$ at the bottom and $I(x,z;\tau\!=\!0)$ on the left of each panel. The V-wave is \textit{not} propagation-invariant, and the rates of diffractive spatial spreading and dispersive temporal spreading are equal.}
\label{fig:concept}
\end{figure*}

\section{Theory of V-waves}

We first briefly review the formulation of ST wave packets \cite{Yessenov19PRA,Yessenov19OE}. The angular spectrum for the slowly varying envelope $\psi(x,z;t)$ of a scalar wave packet $E(x,z;t)\!=\!e^{i(k_{\mathrm{o}}z-\omega_{\mathrm{o}}t)}\psi(x,z;t)$ can be expressed as:
\begin{equation}
\psi(x,z;t)=\iint\!dk_{x}d\Omega\,\widetilde{\psi}(k_{x},\Omega) e^{i\{k_{x}x+(k_{z}-k_{\mathrm{o}})z-\Omega t\}},
\end{equation}
where the spatio-temporal spectrum $\widetilde{\psi}(k_x,\Omega)$ is the 2D Fourier transform of $\psi(x,0;t)$, $\Omega\!=\!\omega-\omega_{\mathrm{o}}$ is the temporal frequency with respect to a fixed frequency $\omega_{\mathrm{o}}$, and $k_{\mathrm{o}}\!=\!\tfrac{\omega_{\mathrm{o}}}{c}$ is its associated wave number. For simplicity, we hold the field uniform along the transverse coordinate $y$. The transverse and longitudinal components of the wave vector $k_{x}$ and $k_{z}$ correspond to the transverse coordinate $x$ and the axial propagation coordinate $z$, respectively, with $k_{x}^{2}+k_{z}^{2}\!=\!(\tfrac{\omega}{c})^{2}$, which is represented geometrically by a light-cone [Fig.~\ref{fig:concept}(a)].

In general, the spectral support domain of a ST wave packet is restricted to 1D curves on the light-cone surface \cite{Yessenov19PRA}. For propagation-invariant ST wave packets, the spectra are conic sections at the intersection of the light-cone with planes that are parallel to the $k_{x}$-axis. So-called `baseband' ST wave packets, whereupon spatial frequencies in the vicinity of $k_{x}\!=\!0$ are permissible \cite{Yessenov19PRA}, are associated with the plane $\Omega\!=\!(k_{z}-k_{\mathrm{o}})c\tan{\theta}$; where the spectral tilt angle $\theta$ is measured with respect to the $k_{z}$-axis [Fig.~\ref{fig:concept}(a)]. Consequently,
\begin{equation}
\psi_{\mathrm{ST}}(x,z;t)\!=\!\int\!\!d\Omega\widetilde{\psi}(\Omega)e^{i\{k_{x}x-\Omega(t-\tfrac{z}{\widetilde{v}})\}}\!=\!\psi_{\mathrm{ST}}(x,0;t-\tfrac{z}{\widetilde{v}}),
\end{equation}
where $k_{x}$ and $k_{z}$ are \textit{both} determined by $\Omega$. The wave packet propagates rigidly in free space at a group velocity $\widetilde{v}\!=\!\tfrac{d\omega}{dk_{z}}\!=\!c\tan{\theta}$, which can take on arbitrary values \cite{Salo01JOA,Recami03IEEEJSTQE,Valtna07OC,Zamboni08PRA,Wong17ACSP2,Efremidis17OL,Porras17OL,PorrasPRA18,Kondakci19NC,Bhaduri19Optica}, and becomes luminal $\widetilde{v}\!=\!c$ only when it degenerates into a plane-wave pulse at $\theta\!=\!45^{\circ}$ \cite{Yessenov19PRA}. Propagation-invariance is highlighted in Fig.~\ref{fig:concept}(b) where we plot the spatio-temporal profiles $I_{\mathrm{ST}}(x,z;t)\!=\!|\psi_{\mathrm{ST}}(x,z;t)|^{2}$ for a ST wave packet with $\theta\!\approx\!49^{\circ}$ (superluminal with $\widetilde{v}\!\approx\!1.15c$) at different axial planes $z$. 

The spectral projection of a ST wave packet onto the $(k_{z},\tfrac{\omega}{c})$-plane is a straight line making an angle $\theta$ with respect to the $k_{z}$-axis [Fig.~\ref{fig:concept}(a)], and that onto the $(k_{x},\tfrac{\omega}{c})$-plane is a conic section that can be approximated by a parabola in the small-bandwidth paraxial regime ($\Delta k_{x}\!\ll\!k_{\mathrm{o}}$ and $\Delta\Omega\!\ll\!\omega_{\mathrm{o}}$),
\begin{equation}\label{eq:STcondition}
\frac{\Omega}{\omega_{\mathrm{o}}}=\frac{1}{2(1-\cot{\theta})}\,\,\frac{k_{x}^{2}}{k_{\mathrm{o}}^{2}}.
\end{equation}

\begin{figure}[t!]
\begin{center}
\includegraphics[width=8.6cm]{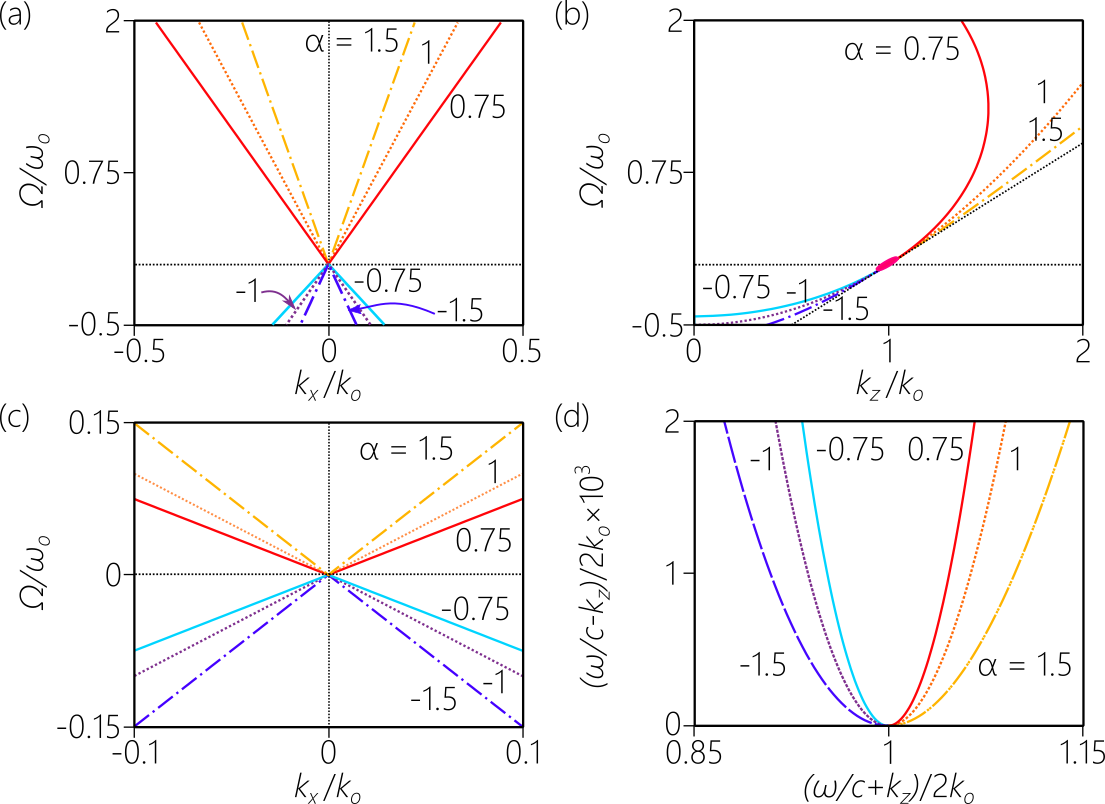} 
\end{center}
\caption{Spectral projections of a V-wave. (a) Spectral projections onto the $(k_{x},\tfrac{\omega}{c})$-plane for V-waves of different values of $\alpha$. The branches of the V-shaped spectrum are continuous through the origin for $\pm\alpha$. (b) Spectral projections onto the $(k_{z},\tfrac{\omega}{c})$-plane of the V-waves from (a). Each curve comprises two segments for $\Omega\!>\!0$ and $\Omega\!<\!0$, corresponding to $\alpha$ and $-\alpha$, respectively, that are tangential to the light-line. (c) Same as (a) but with a restricted range of values for $k_{x}$ and $\Omega$. (d) To highlight the curvature of the spectral projections onto the $(k_{z},\tfrac{\omega}{c})$-plane in (c), the projections are rotated by $45^{\circ}$ so that the light-line coincides with the horizontal axis.}
\label{fig:Theory}
\end{figure}

\begin{figure}[t!]
\begin{center}
\includegraphics[width=8.6cm]{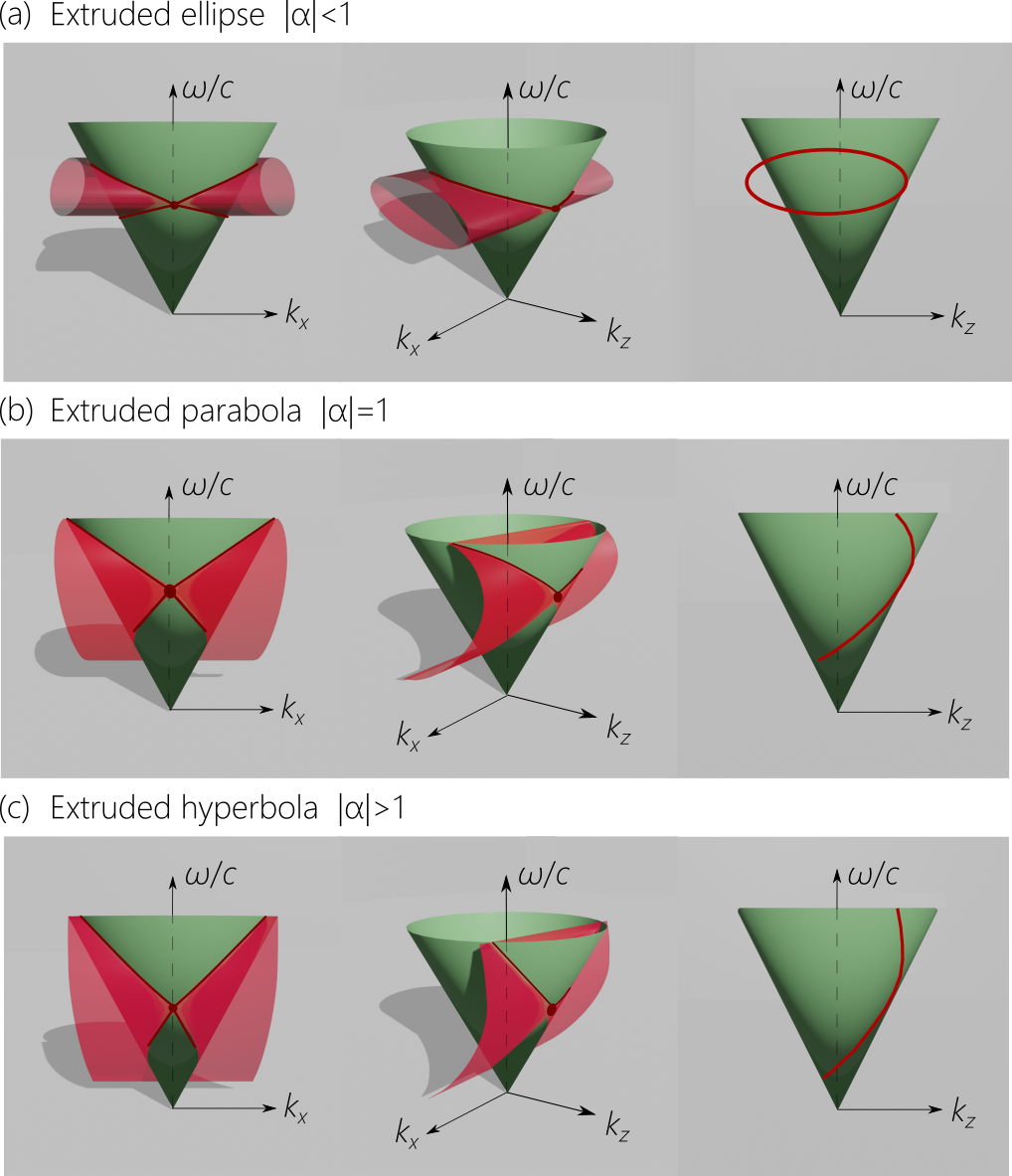} 
\end{center}
\caption{The surface intersecting with the light-cone to yield the spectral support domain of a V-wave is an extruded conic section, whose extrusion axis is parallel to the $k_{x}$-axis. (a) When $|\alpha|\!<\!1$, the surface is an extruded ellipse; (b) when $|\alpha|\!=\!1$, an extruded parabola; and (c) when $|\alpha|\!>\!1$, an extruded hyperbola. In all cases, the extruded conic section is tangential to the light-cone at the point $(k_{x},k_{z},\tfrac{\omega}{c})\!=\!(0,k_{\mathrm{o}},k_{\mathrm{o}})$.}
\label{fig:ExtrudedConicSections}
\end{figure}

The spectral signature of a V-wave is its linear relationship between $\Omega$ and $k_{x}$ [Fig.~\ref{fig:concept}(c)], leading to a V-shaped spatio-temporal spectrum [Fig.~\ref{fig:Theory}(a,c)],
\begin{equation}\label{eq:Vcondition}
\frac{\Omega}{\omega_{\mathrm{o}}}=\alpha\,\,\frac{|k_{x}|}{k_{\mathrm{o}}},
\end{equation}
where $\alpha$ is the dimensionless slope in the $(k_{x},\tfrac{\omega}{c})$-plane. An upright V-shaped spectrum is associated with positive $\alpha$, which is inverted for negative $\alpha$. The corresponding spectral projection onto the $(k_{z},\tfrac{\omega}{c})$-plane is curved [Fig.~\ref{fig:concept}(c) and Fig.~\ref{fig:Theory}(b)]. Two observations are clear from Fig.~\ref{fig:Theory}(c,d): the branches of the V-shaped spectra are continuous through the origin for $\pm\alpha$; and the curved spectra for $\pm\alpha$ in the $(k_{z},\tfrac{\omega}{c})$-plane are continuous through $\Omega\!=\!0$, and are tangential to the light-line. Therefore, V-wave spectra corresponding to $\pm\alpha$ are two branches ($\Omega\!>\!0$ and $\Omega\!<\!0$) of the \textit{same} spectral support domain. 

In contrast to propagation-invariant ST wave packets, the spectral support domain for a V-wave does \textit{not} result from the intersection of the light-cone with a plane. It can be shown that the spectral projection onto the $(k_{z},\tfrac{\omega}{c})$-plane is a conic section,
\begin{equation}\label{Eq:ExtrudedConicSection}
\frac{(\alpha^{2}-1)^{2}}{\alpha^{2}}\left(\frac{\omega/c}{k_{\mathrm{o}}}+\frac{1}{\alpha^{2}-1}\right)^{2}-(\alpha^{2}-1)\frac{k_{z}^{2}}{k_{\mathrm{o}}^{2}}=1,
\end{equation}
which is an ellipse when $|\alpha|\!<\!1$, a hyperbola when $|\alpha|\!>\!1$, and a parabola $2k_{\mathrm{o}}\tfrac{\omega}{c}\!=\!k_{z}^{2}+k_{\mathrm{o}}^{2}$ when $|\alpha|\!=\!1$. Therefore, the spectral support domain for V-waves results from the intersection of the light-cone with \textit{an extruded conic section}, whose extrusion axis is parallel to the $k_{x}$-axis [Fig.~\ref{fig:ExtrudedConicSections}]. The surface is an extruded ellipse when $|\alpha|\!<\!1$ [Fig.~\ref{fig:ExtrudedConicSections}(a)], an extruded parabola when $|\alpha|\!=\!1$ [Fig.~\ref{fig:ExtrudedConicSections}(b)], and an extruded branch of a hyperbola when $|\alpha|\!>\!1$ [Fig.~\ref{fig:ExtrudedConicSections}(c)]. In all cases, the extruded conic section is \textit{tangential} to the light-cone surface at the point $(k_{x},k_{z},\tfrac{\omega}{c})\!=\!(0,k_{\mathrm{o}},k_{\mathrm{o}})$.

\begin{figure*}[ht!]
\begin{center}
\includegraphics[width=17.8cm]{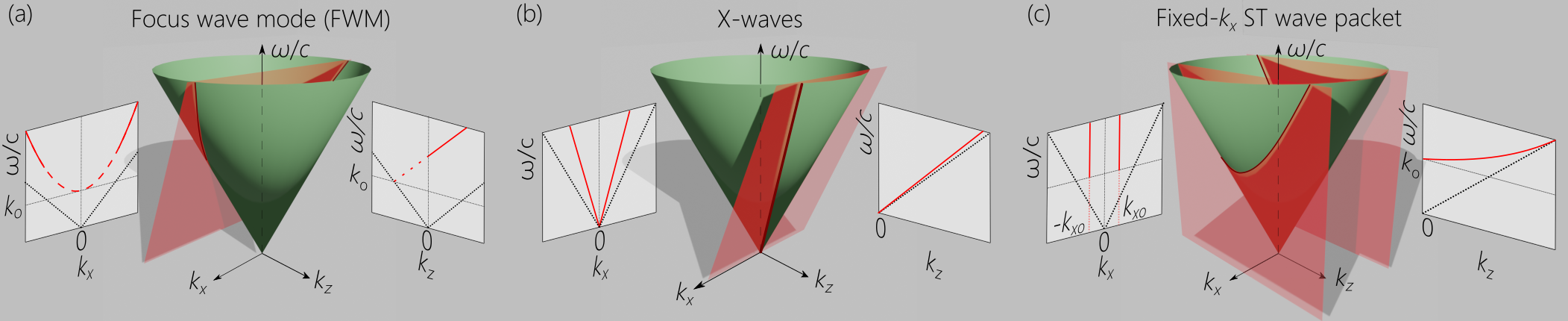} 
\end{center}
\caption{Uniqueness of the attributes of V-waves. (a) Spectral support domain for FWMs: $\widetilde{v}\!=\!c$, GVD is absent, and the projection onto the $(k_{x},\tfrac{\omega}{c})$-plane is a parabola. (b) Spectral support domain for X-waves: $\widetilde{v}\!>\!c$, the projection onto the $(k_{x},\tfrac{\omega}{c})$-plane is V-shaped, but GVD is absent. (c) Spectral support domain for a wave packet with constant $|k_{x}|\!=\!k_{x,\mathrm{o}}$: $\widetilde{v}\!<\!c$, GVD is present, but the projection onto the $(k_{x},\tfrac{\omega}{c})$-plane is not V-shaped. Dotted lines are the light-lines $k_{z}\!=\!\tfrac{\omega}{c}$.}
\label{fig:Uniqueness}
\end{figure*}

\section{Characteristics of V-waves}

Unique characteristics of V-waves arise from the particular form taken by their spatio-temporal spectrum. First, V-waves in free space are luminal $\widetilde{v}\!=\!\frac{d\Omega}{dk_z}\big|_{\Omega=0}\!=\!c$ because the spectral projection onto the $(k_{z},\tfrac{\omega}{c})$-plane is \textit{always} tangential to the light-line. Second, V-waves experience GVD in free space as shown in the axial evolution depicted in Fig.~\ref{fig:concept}(d). Initially at $z\!=\!0$, the profiles for the ST wave packet [Fig.~\ref{fig:concept}(b)] and the V-wave [Fig.~\ref{fig:concept}(d)] are similar. As the V-wave propagates, the on-axis temporal profile $\psi(0,z;t)$ undergoes dispersive broadening and the spatial profile $\psi(x,0;t)$ undergoes diffractive broadening. The GVD parameter for a V-wave is:
\begin{equation}\label{Eq:GVDparameterVwave}
k_{2}=\frac{d^{2}k_{z}}{d\Omega^{2}}\big|_{\Omega=0}=-\frac{1}{\alpha^{2}c\omega_{\mathrm{o}}};
\end{equation}
i.e., the GVD in free space is \textit{anomalous}, and its magnitude can be -- in principle -- tuned to arbitrarily large values. Crucially, unlike the GVD resulting from chromatic dispersion in a resonant optical material or photonic structure, there are no optical losses involved, wavelength restrictions \cite{Yessenov20OSAC}, nor bandwidth limitations \cite{Kondakci18OE}.

Third, the diffraction and dispersion lengths of a V-wave are intrinsically equal. To elucidate this fact, we consider the axial wave number $k_z$ for two different field configurations. First, for a pulsed beam in the paraxial regime propagating in free space $k_{z}\!\approx\! k_{\mathrm{o}}+\frac{\Omega}{c}-\frac{k_{x}^{2}}{2k_{\mathrm{o}}}$, where diffraction is induced by the $k_{x}^{2}$ term. Second, for a plane-wave pulse propagating in a dispersive medium $k_{z}\!\approx\! nk_{\mathrm{o}}+\frac{\Omega}{\widetilde{v}}+\frac{1}{2}k_{2}\Omega^{2}$, where GVD is induced by the $\Omega^2$ term. If we equate the quadratic terms $k_{2}\Omega^{2}\!=\!-\frac{k_{x}^{2}}{k_{\mathrm{o}}}$, we find that imposing a linear relationship between the values of $k_{x}$ and $\Omega$ (that is, a V-wave) achieves our goal $\Omega\!=\!\Omega(k_{x})\!=\!\alpha c|k_{x}|$, with the added constraint $\alpha c\!=\!1/\sqrt{-k_{2}k_{\mathrm{o}}}$, where only negative-valued $k_{2}$ is allowed. Consequently, the spectral phase has a `Janus' structure whereby the axial wave number can be written in terms of either the spatial or temporal frequency:
\begin{equation}
k_{z}= k_{\mathrm{o}}+\frac{\Omega}{c}-\frac{k_{x}^{2}}{2k_{\mathrm{o}}}=k_{\mathrm{o}}+\frac{\Omega}{c}+\frac{1}{2}k_{2}\Omega^{2},
\end{equation}
where we assume propagation in free space so that $\widetilde{v}\!=\!c$. As a result, the spectral phase responsible for diffraction ($-\frac{k_x^2}{2k_o}$) and that for dispersion ($\frac{1}{2}k_2\Omega^2$) are linked, so that they progress at intrinsically equal rates; see Fig.~\ref{fig:concept}(d).

The V-wave envelope travels at a group velocity $\widetilde{v}\!=\!c$ and takes the form:
\begin{equation}
\psi_{\mathrm{V}}(x,z;t)=\int\!dk_x\,\widetilde{\psi}(k_{x}) e^{-i\frac{k_{x}^{2}}{2k_{\mathrm{o}}}z}e^{-i\alpha c|k_{x}|(t-z/c)}e^{ik_{x}x}.
\end{equation}
Indeed, this equation predicts dispersive on-axis pulse broadening associated with a medium having a GVD parameter $k_{2}\!=\!-\tfrac{1}{\alpha^{2}c\omega_{\mathrm{o}}}$. The on-axis $x\!=\!0$ pulse profile has the form:
\begin{equation}
\psi_{\mathrm{V}}(0,z;t)=\int d\Omega\widetilde{\psi}(\Omega)e^{i\tfrac{1}{2}k_{2}\Omega^{2}z}e^{-i\Omega(t-z/c)},
\end{equation}
corresponding to dispersive temporal broadening, and the spatial profile at the pulse center is:
\begin{equation}
\psi_{\mathrm{V}}(x,z;z/c)=\int dk_{x}\widetilde{\psi}(k_{x})e^{i\tfrac{k_{x}^{2}}{2k_{\mathrm{o}}}z}e^{ik_{x}x},
\end{equation}

Finally, we note that the time-averaged intensity $I_{\mathrm{V}}(x,z)\!=\!\int\!dt|\psi_{\mathrm{V}}(x,z;t)|^{2}$ shows no axial evolution,
\begin{equation}
I_{\mathrm{V}}(x,z)=\int\!dk_{x}|\widetilde{\psi}(k_{x})|^{2}+\int\!dk_{x}\widetilde{\psi}(k_{x})\widetilde{\psi}^{*}(-k_{x})e^{i2k_{x}x},
\end{equation}
despite the evolving underlying spatio-temporal profile $I_{\mathrm{V}}(x,z;t)$.


We discuss here the status of V-waves as the unique luminal wave packets that allow for tunable GVD, while maintaining a linear relationship between temporal and spatial frequencies. 

We have compared V-waves above to baseband propagation-invariant ST wave packets. Two families, sideband ST wave packets and X-waves, complete the enumeration of propagation-invariant wave packets \cite{Yessenov19PRA}. The spectral support domain for \textit{sideband} ST wave packets lies at the intersection of the light-cone with the plane $\Omega\!=\!(k_{z}+k_{\mathrm{o}})c\tan{\theta}$ \cite{Yessenov19PRA}. The group velocity is still $\widetilde{v}\!=\!c\tan{\theta}$ with $45^{\circ}\!\leq\!\theta\!<\!90^{\circ}$ (and only FWMs are luminal $\widetilde{v}\!=\!c$ at $\theta\!=\!45^{\circ}$) and the wave packets are GVD-free; see Fig.~\ref{fig:Uniqueness}(a). The spectral support domain for \textit{X-waves} is at the intersection of the light-cone with the plane $\omega\!=\!k_{z}c\tan{\theta}$ [Fig.~\ref{fig:Uniqueness}(b)]. The spatio-temporal spectrum is V-shaped, but its apex corresponding to $k_{x}\!=\!0$ is at $\omega\!=\!0$, and the group velocity is $\widetilde{v}\!=\!c\tan{\theta}$ with $45^{\circ}\!<\!\theta\!<\!90^{\circ}$, so that $\widetilde{v}\!>\!c$. In contrast to V-waves, the spectral projection onto the $(k_{z},\tfrac{\omega}{c})$-plane is a straight line, indicating absence of GVD. Therefore the defining characteristics of V-waves are unique amongst all other ST wave packets.

A previously proposed strategy for achieving GVD in free space is to restrict a ST wave packet to a fixed transverse wave number \cite{Liu98JMO,Hu02JOSAA,Lu03JOSAA}. For two transverse dimensions, this corresponds to a pulsed Bessel beam, and for one transverse dimension a pulsed cosine wave. The spectral support domain for such a wave packet lies at the intersection of the light-cone with vertical planes at $k_{x}\!=\!\pm k_{x,\mathrm{o}}$ [Fig.~\ref{fig:Uniqueness}(c)]. The group velocity is subluminal $\widetilde{v}\!=\!c\cos{\varphi}\!<\!c$, where $\cos{\varphi}\!=\!\sqrt{1-(k_{x,\mathrm{o}}/k_{\mathrm{o}})^{2}}$, and the anomalous GVD parameter is $k_{2}\!=\!-\tfrac{1}{k_{\mathrm{o}}c^{2}}\tan^{2}{\varphi}\sec{\varphi}$. Tuning the GVD necessitates varying $k_{x,\mathrm{o}}$, and therefore changing the transverse spatial scale of the wave packet. To the best of our knowledge, this strategy has \textit{not} been realized experimentally to date.

A V-wave has a GVD parameter of $k_{2}\!\approx\!-\tfrac{2000}{\alpha^{2}}$~fs$^{2}$/mm at $\lambda_{\mathrm{o}}\!\sim\!1$~$\mu$m. Tuning the value of $\alpha$ can thus lead to dramatic changes in GVD. For example, choosing $\alpha\!=\!1$ leads to $k_{2}\!\approx\!-2000$~fs$^{2}$/mm, which approaches that of ZnSe ($k_{2}\!\approx\!1000$~fs$^{2}$/mm at $\lambda\!\sim\!800$~nm) and is much larger than that of silica ($k_{2}\!\approx\!-26$~fs$^2$/mm at $\lambda\!\sim\!1.5$~$\mu$m). Values of $\alpha\!<\!1$ are more convenient to implement experimentally, and V-waves thus readily enable extremely high GVD values. On the other hand, the GVD produced by the constant-$k_{x}$ wave packet is $k_{2}\!\approx\!-2000\tan^{2}{\varphi}\sec{\varphi}$~fs$^{2}$/mm at a wavelength $\lambda_{\mathrm{o}}\!\sim\!1$~$\mu$m. Achieving large GVD requires large $\varphi$ approaching the non-paraxial regime. For example, whereas $\varphi\!\approx\!6.5^{\circ}$ yields $k_{2}\!\approx\!-28.3$~fs$^{2}$/mm rivaling fused silica, reaching $k_{2}$ for ZnSe requires $\varphi\!\approx\!35^{\circ}$. In comparison, V-waves can provide larger GVD at smaller numerical apertures.

\section{Experiment}

\begin{figure}[t!]
\begin{center}
\includegraphics[width=8.6cm]{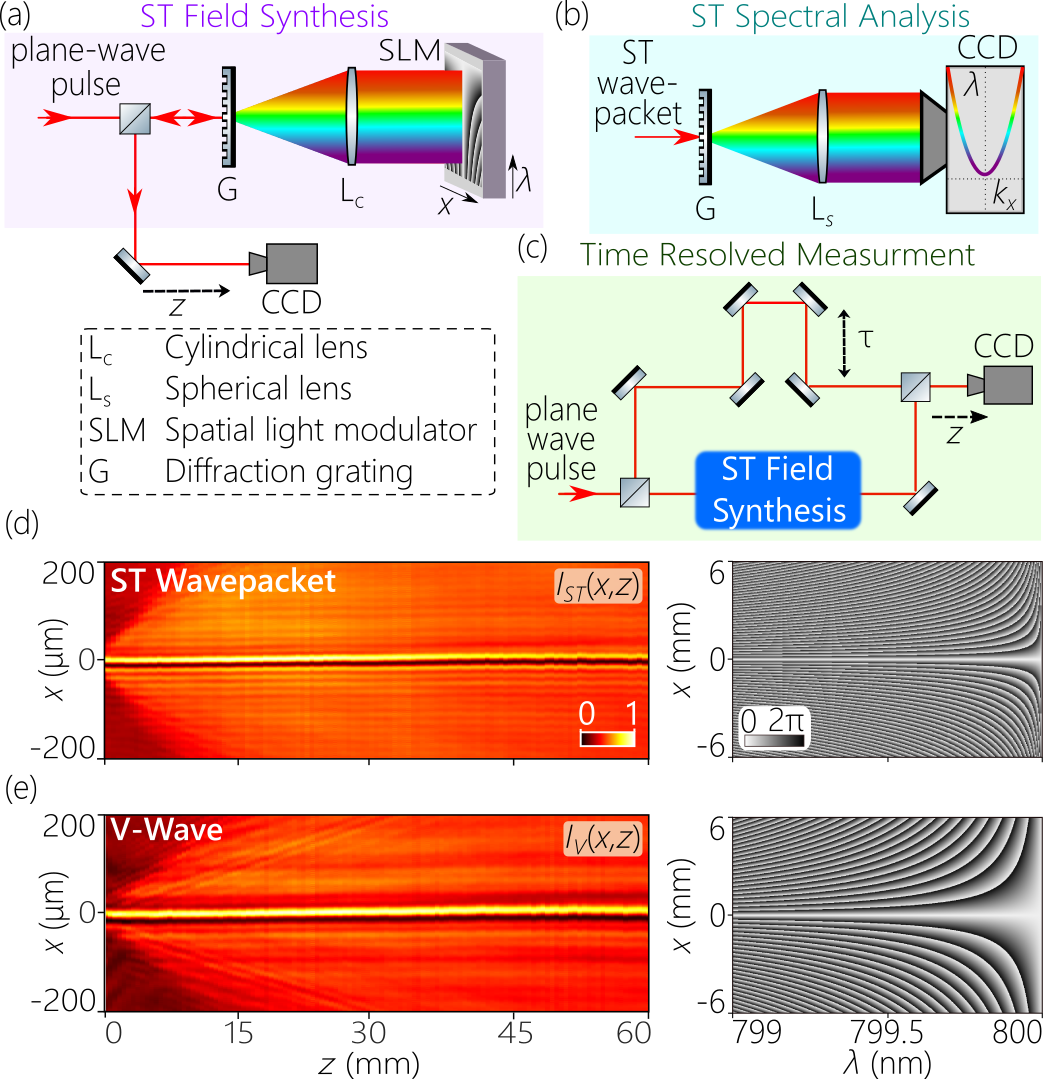} 
\end{center}
\caption{Schematics of the experimental arrangements (a) for synthesizing ST wave packets and V-waves, (b) measuring the spatio-temporal spectrum $|\widetilde{\psi}(k_{x},\lambda)|^{2}$, and (c) reconstructing the spatio-temporal profile $I(x,z;\tau)$ at a fixed plane $z$. (d) Measured time-averaged intensity $I_{\mathrm{ST}}(x,z)$ for a ST wave packet with $\theta\!=\!49^{\circ}$ and $\Delta\lambda\!\approx\!2$~nm. On right we plot a section of the phase distribution $\Phi$ imparted by the SLM. (e) Same as (d) for a V-wave having $\alpha\!=\!0.1$ and $\Delta\lambda\!\approx\!2$~nm. Note the different spatial scales used along $x$ and $z$ in (d) and (e).}
\label{fig:Setup}
\end{figure}

To study the behavior of V-waves experimentally, we make use of the optical system developed in Refs.~\cite{Yessenov19PRA,Kondakci19OL,Kondakci17NP}, which allows for the synthesis of ST wave packets of arbitrary spatio-temporal spectral structure; see Fig.~\ref{fig:Setup}(a). Femtosecond pulses from a mode-locked Ti:sapphire laser (Tsunami; Spectra Physics) at a wavelength of $\lambda_{\mathrm{o}}\!\sim800$~nm are incident on a diffraction grating (Newport 10HG1200-800-1) having 1200~lines/mm and an area of $25\times25$~mm$^{2}$. The first diffraction order is collimated with a cylindrical lens in a $2f$ configuration before impinging on a 2D reflective, phase-only spatial light modulator (SLM; Hamamatsu X10468-02), whereupon each wavelength is imparted a phase distribution $\Phi$ along the direction orthogonal to the spread spectrum. This phase distribution is designed to assign a spatial frequency $k_{x}$ to each wavelength $\lambda$ in accordance with Eq.~\ref{eq:STcondition} for propagation-invariant ST wave packets, and Eq.~\ref{eq:Vcondition} for V-waves. The phase-modulated wave front is retro-reflected to the grating, whereupon the wave packet is reconstituted.

\begin{figure*}[t!]
\begin{center}
\includegraphics[width=17.6cm]{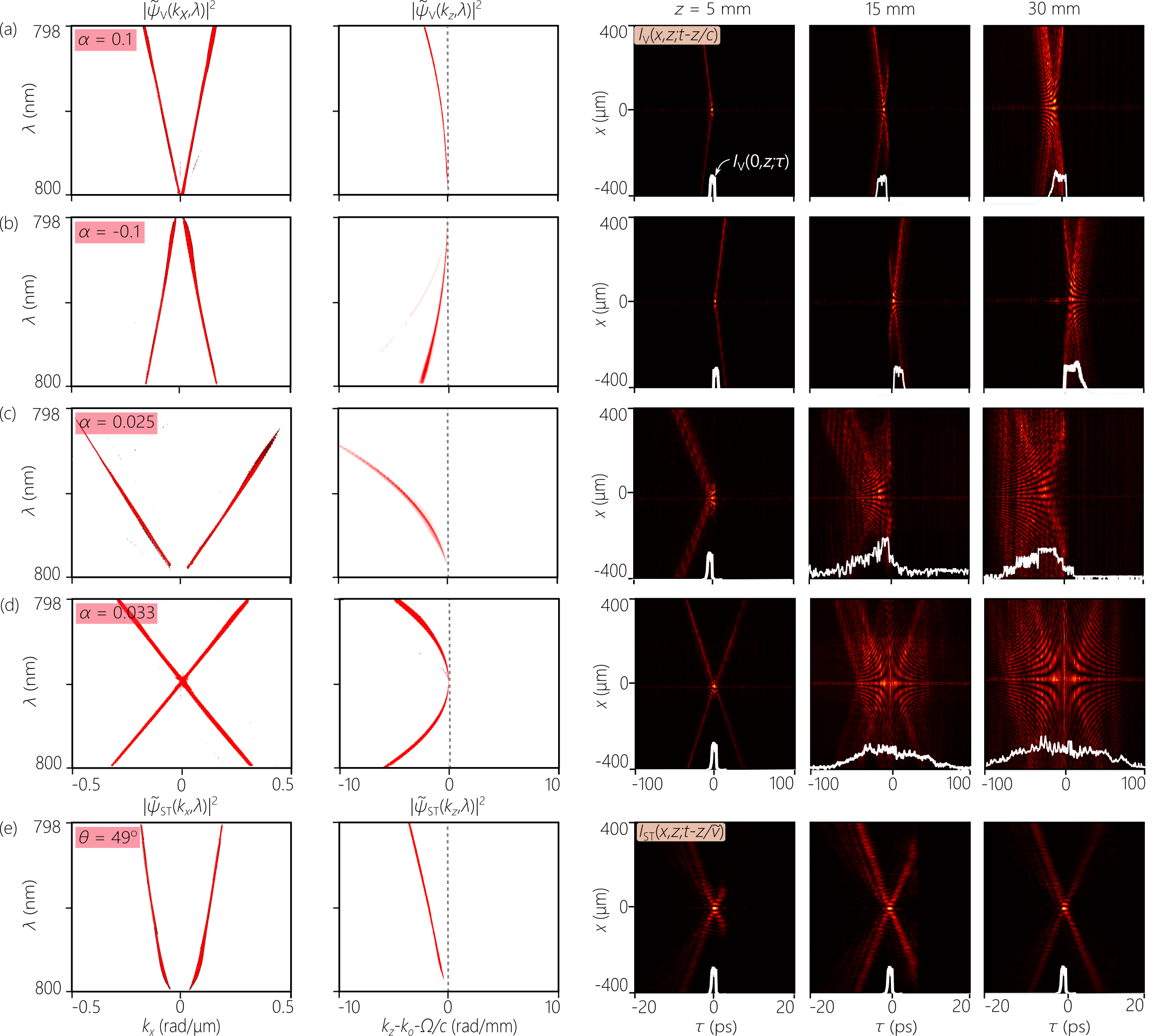} 
\end{center}
\caption{Experimental demonstration of V-waves. The first and second columns are the spectral projection onto the $(k_{x},\lambda)$ and $(k_{z},\lambda)$ planes. The last three columns are spatio-temporal intensity profiles $I(x,z;\tau)$ measured at $z\!=\!5$, 15, and 30~mm, and the white curve in each panel is the temporal profile $I(x\!=\!0,\tau;z)$. (a) A V-wave with $\lambda_{\mathrm{o}}\!=\!800$~nm, $\alpha\!=\!-0.1$; (b) $\alpha\!=\!0.1$; and (c) $\alpha\!=\!0.025$. (d) A V-wave with $\lambda_{\mathrm{o}}\!=\!799$~nm, $\alpha\!=\!\pm0.033$. In (a-d) measurements are carried out in a frame traveling at $c$, $\tau\!=\!t-z/c$. (e) A propagation-invariant ST wave packet with $\Delta k_{x}$, $\Delta\lambda$, and $\lambda_{\mathrm{o}}$ equal to those in (a);  the spectral tilt angle is $\theta\!=\!49^{\circ}$, and measurements are carried out in a frame traveling at $\widetilde{v}\!=\!c\tan{\theta}$, $\tau\!=\!t-z/\widetilde{v}$.}
\label{fig:Measurements}
\end{figure*}

To characterize the synthesized field, three methodologies are implemented. First, the time-averaged intensity $I(x,z)\!=\!\int\!dt|\psi(x,z;t)|^{2}$ is recorded with an axially scanned CCD camera; Fig.~\ref{fig:Setup}(a). Second, the spatio-temporal spectrum $|\widetilde{\psi}(k_{x},\lambda)|^{2}$ is acquired by resolving the wavelengths with a diffraction grating and performing a spatial Fourier transform using a lens, from which we extract the spectral projection onto the $(k_{z},\tfrac{\omega}{c})$-plane and thus confirm $\widetilde{v}$ and $k_{2}$; Fig.~\ref{fig:Setup}(b). Third, a two-path interformetric arrangement reconstructs the spatio-temporal profile of the ST wave packet $I(x,z;\tau)\!=\!|\psi(x,z;\tau)|^{2}$ at fixed axial planes. The synthesis system from Fig.~\ref{fig:Setup}(a) is placed in one arm, and the original femtosecond laser pulses are directed to a reference arm containing an optical delay $\tau$; Fig.~\ref{fig:Setup}(c). When the wave packet and the reference pulse overlap in space and time, we observe spatially resolved fringes from whose visibility we can reconstruct $I(x,z;\tau)$ \cite{Kondakci19NC,Bhaduri19Optica}. This arrangement facilitates monitoring the axial propagation of the wave packet, and thus directly assessing $\widetilde{v}$ and $k_{2}$.

We plot in Fig.~\ref{fig:Setup}(d) and Fig.~\ref{fig:Setup}(e) the time-averaged intensity $I_{\mathrm{ST}}(x,z)$ for a ST wave packet ($\theta\!\approx\!49^{\circ}$) and $I_{\mathrm{V}}(x,z)$ for a V-wave ($\alpha\!\approx\!0.1$), respectively. As expected, both display clear diffraction-free propagation. We select $\theta\!=\!49^{\circ}$ to ensure the same spatial bandwidth $\Delta k_{x}\!=\!0.2$~rad/$\mu$m$^2$ \textit{and} temporal bandwidth $\Delta\lambda\!=\!2$~nm as the V-waves with $\alpha\!=\!\pm0.1$, as determined by $\alpha^{2}\!=\!\tfrac{\Delta\lambda}{2\lambda_{\mathrm{o}} |1-\cot{\theta}|}$

We present our spectral and time-resolved measurements for V-waves in Fig.~\ref{fig:Measurements}. For each wave packet we plot three quantities. First, we plot in the first column of Fig.~\ref{fig:Measurements} the spectral intensity $|\widetilde{\psi}(k_{x},\lambda)|^{2}$ projected onto the $(k_{x},\lambda)$-plane to verify the V-shaped spatio-temporal spectrum. Second, we extract from $|\widetilde{\psi}_{\mathrm{V}}(k_{x},\lambda)|^{2}$ the spectral projection onto the $(k_{z},\lambda)$-plane, $|\widetilde{\psi}_{\mathrm{V}}(k_{z},\lambda)|^{2}$, and plot it in the second column of Fig.~\ref{fig:Measurements} in terms of $k_{z}-k_{\mathrm{o}}-\tfrac{\Omega}{c}$ to isolate any nonlinear contributions to the spectral projection. Third, we plot the spatio-temporal profiles $I_{\mathrm{V}}(x,z;\tau)$ at three axial plane ($z\!=\!5$, 15, and 30~mm) in a moving frame traveling at $c$ ($\tau\!=\!t-z/c$) to delineate the propagation dynamics. The wave packets have the same temporal bandwidth $\Delta\lambda\!\approx\!2$~nm, so that the initial on-axis profile $I_{\mathrm{V}}(0,0;\tau)$ are similar.

We plot the results for V-waves with $\alpha\!=\!\pm0.1$ in Fig.~\ref{fig:Measurements}(a,b) to highlight the impact of the sign of $\alpha$. Note that the spectral projection $|\widetilde{\psi}_{\mathrm{V}}(k_{x},\lambda)|^{2}$ is V-shaped and $|\widetilde{\psi}_{\mathrm{V}}(k_{z}-k_{\mathrm{o}}-\tfrac{\Omega}{c},\lambda)|^{2}$ is purely quadratic (one half of a parabola), and both are flipped along the $\lambda$-axis after switching the sign of $\alpha$. The spatio-temporal profiles undergo rapid axial evolution, and the on-axis temporal spreading is consistent with a GVD parameter of $\omega_{\mathrm{o}}ck_{2}\!=\!-\tfrac{1}{\alpha^{2}}\!=\!-100$, which is $\sim4$ order-of-magnitude larger than that of silica (at $\lambda_{\mathrm{o}}\!=\!1.5$~$\mu$m). Changing the sign of $\alpha$ does not affect $k_{2}$, but it does switch the direction of the asymmetric pulse spreading. 

Next we plot measurements for a V-wave with $\alpha\!=\!0.25$ to isolate the effect of the magnitude of $\alpha$. The V-shaped spectral projection $|\widetilde{\psi}_{\mathrm{V}}(k_{x},\lambda)|^{2}$ has a larger opening angle, and the curvature of the parabolic spectral projection $|\widetilde{\psi}_{\mathrm{V}}(k_{z}-k_{\mathrm{o}}-\tfrac{\Omega}{c},\lambda)|^{2}$ is larger, thus indicating a significantly larger GVD parameter -- as confirmed by the on-axis dispersive broadening of the temporal profile. In Fig.~\ref{fig:Measurements}(d) we show a wave-packet synthesized with an X-shaped spectral spectral projection $|\widetilde{\psi}_{\mathrm{V}}(k_{x},\lambda)|^{2}$ that combines V-waves with $\alpha\!=\!0.033$ and $\alpha\!=\!-0.033$. The spectral projection $|\widetilde{\psi}_{\mathrm{V}}(k_{z}-k_{\mathrm{o}}-\tfrac{\Omega}{c},\lambda)|^{2}$ shows both sides of a parabola, and the axial evolution of the spatio-temporal profile is now symmetrized around $\tau\!=\!0$.

For comparison, we plot in Fig.~\ref{fig:Measurements}(e) measurements for a propagation-invariant ST wave packet with $\theta\!\approx\!49^{\circ}$. In contrast to V-waves, the spectral projection $|\widetilde{\psi}_{\mathrm{ST}}(k_{x},\lambda)|^{2}$ is parabolic, whereas $|\widetilde{\psi}_{\mathrm{ST}}(k_{z}-k_{\mathrm{o}}-\tfrac{\Omega}{c},\lambda)|^{2}$ is a straight line, indicating the absence of GVD. The spatio-temporal intensity profiles $I_{\mathrm{ST}}(x,z;\tau)$ confirm the propagation-invariance of this ST wave packet, which travels at a group velocity of $\widetilde{v}\!=\!c\tan{\theta}\!\approx\!1.15c$.

\begin{figure}[t!]
\begin{center}
\includegraphics[width=8.6cm]{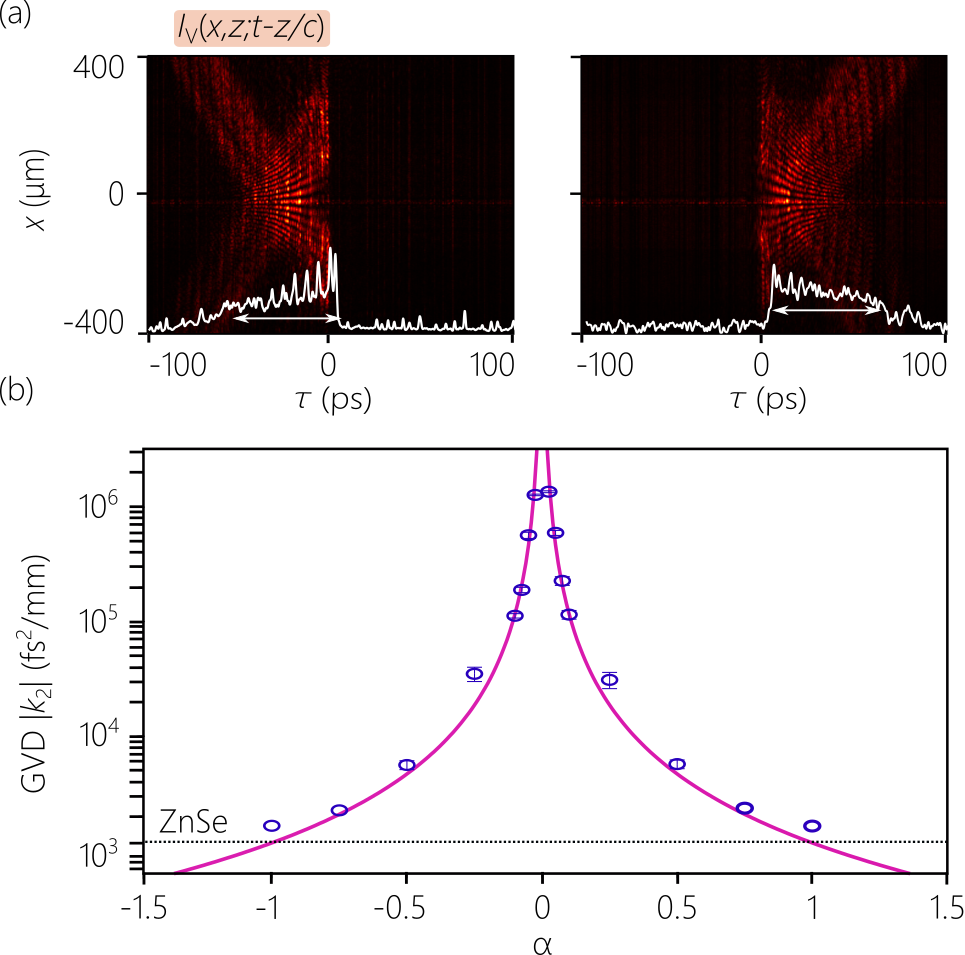} 
\end{center}
\caption{(a) The spatio-temporal profiles for V-waves with $\alpha\!=\!\pm0.025$ at $z\!=\!10$~mm; other parameters similar to Fig.~\ref{fig:Measurements}(a-c). The white curves are $I_{\mathrm{V}}(0,z;\tau)$, which is used to determine $\Delta\tau$. (b) The GVD parameter $k_{2}$ while varying $\alpha$. The circles are data and the curve is the theoretical expectation (Eq.~\ref{Eq:GVDparameterVwave}).}
\label{fig:GVDalpha}
\end{figure}

Finally, to confirm the predicted GVD parameter for V-waves, we measure the full width at $1/e$-the-maximum $\Delta\tau$ of the on-axis temporal intensity profile $I_{\mathrm{V}}(0,z;\tau)$ [Fig.~\ref{fig:GVDalpha}(a)] at $z\!=\!10$, 25, and 50~mm to estimate the GVD parameter $k_{2}\!=\!\Delta\tau/z\Delta\omega$. Measurements of $k_{2}$ while varying $\alpha$ are plotted in Fig.~\ref{fig:GVDalpha}(b) and are in good agreement with the theoretical expectation over three orders-of-magnitude of $k_{2}$.

\section{Discussion and Conclusions}

The new ST wave packet we have introduced here under the moniker `V-wave' is characterized by several unique characteristics, which we summarize as follows:
\begin{enumerate}
    \item The temporal frequency $\Omega$ in a V-wave is linearly related to the \textit{transverse} wave number $k_{x}$, in contrast to propagation-invariant ST wave packets where $\Omega$ is linearly related to the \textit{axial} wave number $k_{z}$ (with the exception of X-waves where $\Omega$ is related linearly to both $k_{x}$ and $k_{z}$).
    \item V-waves in free space are luminal $\widetilde{v}\!=\!c$, in contrast to all propagation-invariant ST wave packets that are not (with the exception of FWMs).
    \item V-waves experience anomalous GVD in free space, and are thus not propagation-invariant. The magnitude of the GVD can be readily tuned to extremely large values that are not available in common optical materials or photonic devices.
    \item The most striking feature of V-waves is that their diffraction and dispersion lengths are intrinsically equal by virtue of the spatio-temporal structure of the field itself.
\end{enumerate}

We recently confirmed a consequence of the latter property in the context of demonstrating the Talbot effect in space and time simultaneously. A periodic profile along $x$ as required for the spatial Talbot effect is achieved by periodically sampling the spatial spectrum along $k_{x}$ \cite{Hall21arxiv}. Only a linear relationship between $k_{x}$ and $\omega$ in the spectrum of the ST field guarantees that $\omega$ is simultaneously sampled periodically, leading to a pulse-train structure in the time domain. The intrinsically equal diffraction and dispersion lengths that is a consequence of the underlying V-shaped spatio-temporal spectrum guarantees equal spatial and temporal Talbot axial self-imaging lengths, which has led to the first observation of the space-time Talbot effect \cite{Hall21arxiv}.

In general, previous work in the area of ST wave packets has focused on their propagation invariance. However, it is now known that controllable axial evolution of an isolated parameter can be realized while holding the other features invariant. For example, the group velocity can evolve axially to produce accelerating or decelerating wave packets \cite{Yessenov20PRL2}, or the on-axis spectrum can evolve axially in a prescribed manner \cite{Motz21PRA}. Recently, it was demonstrated that a propagation-invariant ST wave packet can be accompanied by another wave component that focused abruptly at a prescribed axial location \cite{Wong20AS}. Here we have shown that the on-axis pulse profile can undergo dispersive broadening in time as a result of arbitrary-valued GVD introduced into the field structure.

In conclusion, we have examined theoretically and confirmed experimentally a new family of ST wave packets that we have called V-waves, which are characterized by a V-shaped spatio-temporal spectrum. The linear relationship between temporal frequencies $\omega$ and spatial frequencies $k_{x}$ give rise to a nonlinear relationship between $\omega$ and the axial wave numbers $k_{z}$. Consequently, V-waves can be endowed with anomalous GVD of tunable magnitude in free space. Using picosecond laser pulses, we have realized GVD parameters in the range from $\sim10^{3}$ to $\sim10^{6}$~fs$^2$/mm, all the while maintaining the group velocity fixed at $\widetilde{v}\!=\!c$. This work paves the way for GVD management of ST wave packets in dispersive media, and suggests potential applications in nonlinear optics (including pulse compression and group-velocity matching of different-wavelength pulses).

\section*{Acknowledgments}
We thank H. E. Kondakci for useful discussions. This work was funded by the U.S. Office of Naval Research (ONR) under contracts N00014-17-1-2458 and N00014-19-1-2192.


\bibliography{diffraction}

\end{document}